\documentclass[secnumarabic,amsmath,amssymb, nobibnotes]{revtex4-1}
\usepackage{bmpsize}
\usepackage{graphicx}% Include figure files
\usepackage{dcolumn}% Align table columns on decimal point
\usepackage{bm}% bold math
\usepackage{physics}% physics
\usepackage{braket} % braket
\usepackage{hyperref}
\usepackage{alphabeta}
\usepackage{amsmath}
\usepackage{subfigure}
\usepackage[font=small,labelfont=bf]{caption}
\usepackage{setspace} 
\usepackage{lipsum} % one column text in two column document
\usepackage{soul}
\usepackage{ulem}

\usepackage{booktabs}
\usepackage{alphabeta}
\usepackage{xcolor}

\usepackage{etoolbox}
\patchcmd{\section}
  {\centering}
  {\raggedright}
  {}
  {}
\patchcmd{\subsection}
  {\centering}
  {\raggedright}
  {}
  {}
\patchcmd{\subsubsection}
  {\centering}
  {\raggedright}
  {}
  {}
\usepackage[english]{babel}
\usepackage{blindtext}
\usepackage{verbatim}

\begin{document}

\title{Spin-spin interactions in solids from mixed all-electron and pseudopotential calculations\textemdash a path to screening materials for spin qubits}

\author{Krishnendu Ghosh$^1$}
\thanks{Current affiliation: Computational and Modeling Technology, Intel Corporation, Hillsboro OR 97124}
\author{He Ma$^2$$^,$$^3$}
\author{Mykyta Onizhuk$^2$$^,$$^3$}
\author{Vikram Gavini$^1$$^,$$^4$}
\thanks{Corresponding author:\href{vikramg@umich.edu}{vikramg@umich.edu}}
\author{Giulia Galli$^2$$^,$$^3$$^,$$^5$}
\thanks{Corresponding author:\href{gagalli@uchicago.edu}{gagalli@uchicago.edu}}
\affiliation{$^1$Department of Mechanical Engineering, University of Michigan, Ann Arbor MI 48109}
\affiliation{$^2$Pritzker School of Molecular Engineering, University of Chicago, Chicago IL 60637}
\affiliation{$^3$Department of Chemistry, University of Chicago, Chicago IL 60637}
\affiliation{$^4$Department of Materials Science and Engineering, University of Michigan, Ann Arbor MI 48109}
\affiliation{$^5$Materials Science Division, Argonne National Laboratory, Lemont IL 60439}

\date{\today}
\begin{abstract}
\vspace{20pt}

Understanding the quantum dynamics of spin defects and their coherence properties requires an accurate modeling of spin-spin interaction in solids and molecules, for example by using  spin Hamiltonians with parameters obtained from first-principles calculations. We present a real-space approach based on density functional theory for the calculation of spin-Hamiltonian parameters, where only selected atoms are treated at the all-electron level, while the rest of  the system is described with the pseudopotential approximation. Our approach permits calculations for systems containing more than 1000 atoms, as demonstrated for defects in diamond and silicon carbide. We show that only a small number of atoms surrounding the defect needs to be treated at the all-electron level, in order to obtain an overall all-electron accuracy for hyperfine and zero-field splitting tensors.  We also present results for coherence times, computed with the cluster correlation expansion method,  highlighting the importance of accurate spin-Hamiltonian parameters for quantitative predictions of spin dynamics.

\end{abstract}
\maketitle
%\tableofcontents

\section*{Introduction}

Spin-defects in semiconductors are promising quantum bits (qubits) for quantum information technologies including quantum computation, communication and sensing \cite{Weber2010,Anderson2019}. A prime example of spin-defects is the nitrogen-vacancy (NV) center in diamond \cite{Davies1976,Rogers2008,Doherty2011,Maze2011,Goldman2015}, which can be optically initialized and read-out, and possesses millisecond coherence time even at room temperature. In recent years, much effort has been devoted to realizing novel spin-defects in industrially mature host materials with properties similar or superior to those of diamond NV centers. For instance, several promising spin-defects have been identified in silicon carbide, including the divacancy (VV) \cite{Koehl2011,Whiteley2018}, Cr \cite{Son1999,Koehl2017,Diler2019}, and V impurities \cite{Wolfowicz2020}. There is also a growing interest in discovering and designing spin qubits in piezo-electric materials such as aluminum nitride \cite{Seo2016,Seo2017}, in oxides \cite{Morfa2012} and in 2D materials \cite{Ye2019,Yim2020}.

First-principles calculations based on density functional theory (DFT) have played an important role in the discovery and identification of novel spin-defects, in particular in understanding their electronic and thermodynamical properties \cite{Ivady2018,Dreyer2018}. DFT results have been instrumental in interpreting optical and magnetic measurements, and in predicting atomistic and electronic structures of defects yet to be realized experimentally. In addition, specific spin dynamical properties may be investigated with the aid of DFT calculations, using spin Hamiltonians (SH) with parameters obtained from first principles. For systems with a single effective electron spin, e.g. a spin defect in a semiconductor, the leading terms in the spin Hamiltonian are \cite{Schweiger2001, Harriman2013, Abragam2013}:
\begin{widetext}
\begin{equation} \label{Eq. 1}
H = \mu_B \bm{B} \cdot \bm{g} \cdot \bm{S} + \sum_N \gamma_N \bm{B} \cdot \bm{I}_N + \sum_N \bm{S} \cdot \bm{A}_N \cdot \bm{I}_N + \bm{S} \cdot \bm{D} \cdot \bm{S} + \sum_N \bm{I}_N \cdot \bm{P}_N \cdot \bm{I}_N
\end{equation}
\end{widetext}
where $\mu_B$ is Bohr magneton; $\bm{S}$ is the effective electron spin; $\bm{B}$ is the external magnetic field; $\bm{I}_N$ and $\gamma_N$ are the spin and gyromagnetic ratio of the $N^{\text{th}}$ nucleus; $\bm{g}$, $\bm{A}$, $\bm{D}$, and $\bm{P}$ are rank-2 tensors that characterize the strength of electron Zeeman interaction, hyperfine interaction, zero-field splitting and nuclear quadrupole interaction, respectively. Nuclear spin-spin interactions and the chemical shielding effect in nuclear Zeeman interactions are neglected in Eq. \ref{Eq. 1}.

The spin Hamiltonian parameters $\bm{g}$, $\bm{A}$, $\bm{D}$ and $\bm{P}$ can be determined from first-principles electronic structure calculations \cite{Vandewalle1993,Blugel1987,Overhof2004,Bahramy2007,Rayson2008,Bodrog2013,Biktagirov2018,Olsen2002,Sinnecker2006,Neese2005,Reviakine2006,Kossmann2007,Neese2007}, the majority of which are based on plane-wave pseudopotential (PW-PP) approaches. DFT calculations of SH parameters using basis sets different from PW have been proposed (e.g. numerical atomic orbitals \cite{Kadantsev2008}, linearized augmented plane-wave \cite{Schwarz2003}, linear muffin-tin orbitals \cite{Daalderop1996, Overhof2004}, and Gaussian orbitals \cite{Dovesi2018}), but they are often limited to smaller systems than those accessible to PW calculations. In the PW-PP method, pseudopotentials (PP) are used to describe the interaction between valence and core electrons, and single-particle wavefunctions of core electrons in the solid are not explicitly evaluated. All-electron wavefunctions may be reconstructed, for example using the projected augmented wave (PAW) procedure \cite{Blochl1994}, and then used to compute the parameters of the SH. Recently, we proposed and benchmarked a real-space all-electron DFT framework \cite{ghoshPRM2019} using a finite-element (FE) basis sets \cite{DFTFE-CPC-2020} for accurate predictions of SH parameters in molecules and solids, which does not require any reconstruction of core-wavefunctions. This framework allows one to systematically convergence the results of SH parameters with respect to the basis set size and hence to establish robust results to compare with experiments for a chosen level of first-principles theory. However, the method is computationally rather demanding and it only permitted the investigation of systems with tens of atoms. 

Here we propose a computational scheme where only selected atoms are treated at the all-electron (AE) level, while the rest of the system is described within the PP approximation. We show that in order to obtain accurate SH parameters for spin-defects, only a small number of atoms surrounding the defect (of the order of 10) needs to be treated at the AE level. Our approach permits calculations for cells with hundreds of atoms, as shown for the NV center in diamond and the VV in 4H-SiC, for which we used cells with up to 1022 atoms. In addition, using the cluster correlation expansion (CCE) \cite{Yang2008} method, we demonstrate the importance of accurate SH parameters for precise predictions of coherence times of spin defects in semiconductors. 

\begin{figure}[!ht]
  \centering
  \includegraphics[width=6in]{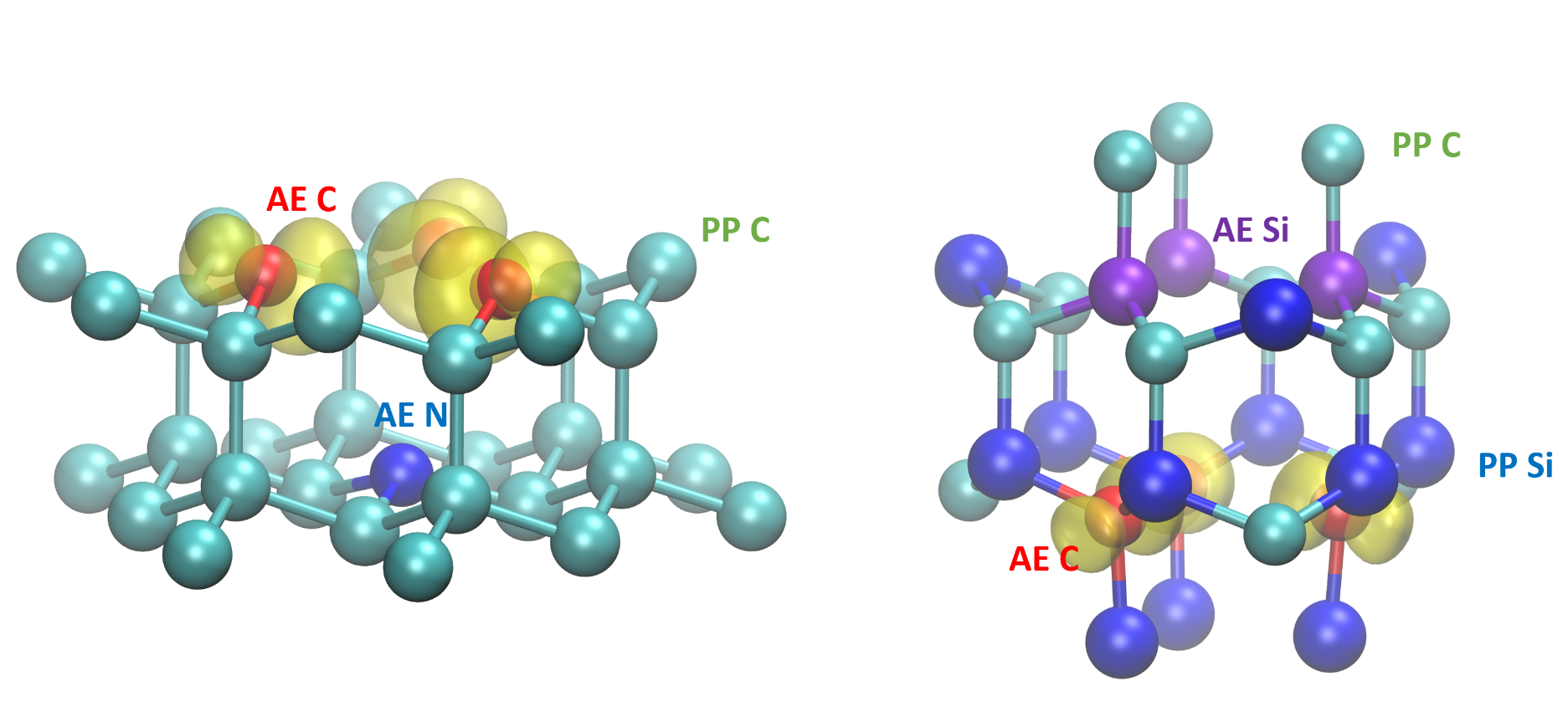}
  \caption{Structures and spin densities of the negatively-charged nitrogen-vacancy center in diamond (left) and the divacancy in 4H-SiC (right). In both systems, the spin density is localized around the carbon atoms with dangling bonds. By only treating a few atoms at the all-electron (AE) level, and the remaining using the pseudopotential (PP) approximation, an accurate prediction of spin Hamiltonian parameters can be achieved.}
  \label{fig_diagram}
\end{figure}

\section*{Results}

We carried out mixed AE-PP DFT calculations of hyperfine constants and zero-field splitting tensors using a finite element (FE) basis and we compared the results to those obtained with PP calculations using plane waves (PW). In particular, we applied the mixed AE-PP approach to the negatively-charged NV center in diamond and neutral VV in 4H-SiC (see Fig. \ref{fig_diagram}). Both defects have spin-triplet ground states. For VV we considered the \textit{kk} configuration, where both the carbon and silicon vacancies are located at quasi-cubic sites. We modeled the diamond and 4H-SiC using cubic and hexagonal supercells, respectively. For consistency, both PW and FE calculations were performed using structures relaxed with PW DFT. All calculations were performed with the Perdew–Burke-Ernzerhof (PBE) exchange-correlational functional~\cite{pbe1996}. The use of the FE basis sets for SH parameter calculations permits an increase in resolution in the core regions of atoms, where wavefunctions exhibit a highly oscillatory behavior, while a coarser resolution suffices in the valence region. Due to its spatial adaptivity, the FE basis set is perfectly suited to carry out mixed AE-PP calculations. We briefly summarize below our strategy to evaluate hyperfine and zero-field splitting tensors, before presenting our results.

\subsection*{Computational framework for hyperfine and zero-field splitting tensors}

The hyperfine $\bm{A}$-tensor is composed of an isotropic part $A^{\text{fc}}$ originating from the Fermi contact of electrons at the nuclei, and an anisotropic part $A^{\text{sd}}$ originating from spin dipolar interactions. $A^{\text{fc}}$ and $A^{\text{sd}}$ can be evaluated using the electron spin density of the system as:

\begin{equation} \label{Eq. 2} 
    A^{\text{fc}} = - \frac{1}{3S} \mu_0 \gamma_e \gamma_N \hbar^2 n_s(\bm{r}_N),
\end{equation}

\begin{widetext}
\begin{equation} \label{Eq. 3}
    A^{\text{sd}}_{ab} = \frac{1}{2S} \frac{\mu_0}{4\pi} \gamma_e \gamma_N \hbar^2 \int \frac{|\bm{r} - \bm{r}_N|^2 \delta_{ab} - 3 (\bm{r} - \bm{r}_N)_a (\bm{r} - \bm{r}_N)_b}{|\bm{r} - \bm{r}_N|^5} n_s(\bm{r}) d\bm{r},
\end{equation}
\end{widetext}
where $n_s(\textbf{r})$ is the electronic spin density in real space, $\bm{r}_N$ is the position of the nucleus of interest, and $(\bm{r} - \bm{r}_N)_a$ is the $a$-direction component of $\bm{r} - \bm{r}_N$. S is the spin quantum number of the system ($S=0$ for a singlet, $ \frac{1}{2}$ for a doublet, etc.), $\gamma_e$ and $\gamma_N$ are gyromagnetic ratios for electron and nuclei, respectively. We note that, while the Fermi contact term requires an accurate value of the electronic spin density exactly {\it at} the nucleus, the spin-dipole term requires an accurate description of the spin density in a region surrounding the nucleus, with the size of the region determined by the compact support of the spin density (i.e. the region where the spin density is non-negligible) and by the decay of the kernel in Eq.~\eqref{Eq. 3}. The spatial adaptivity of the FE basis is important for accurate descriptions of the spin density both in the core and valence regions, thus providing an efficient and systematic means of obtaining converged results, as described in the Methods section.

The spin-spin component of the zero-field splitting tensor $\bm{D}$ is evaluated as the expectation value of the magnetic dipole-dipole operator, $ \frac{ \tilde{r}^2\delta_{ab} - 3\tilde{r}_a \tilde{r}_b }{ r^5 } $, using a Slater determinant built from Kohn-Sham orbitals ~\cite{ Rayson2008, ghoshPRM2019}.  The operator is essentially the $ab$ element of the Hessian of the Green's function, $G(\textbf{r},\textbf{r}^{\ensuremath{'}})=\frac{1}{|\textbf{r}-\textbf{r}^{\ensuremath{'}}|}$; $\tilde{r}$ is a scalar representing $|\textbf{r}-\textbf{r}^{\ensuremath{'}}|$; $\tilde{r}_a$ is the $a$-th component of the vector $\textbf{r}-\textbf{r}^{\ensuremath{'}}$ . A straightforward evaluation of the $\bm{D}$-tensor in real-space involves computing double integrals that are computationally very demanding. In our recent work~\cite{ghoshPRM2019}, we reformulated the evaluation of the $\bm{D}$-tensor in real-space by solving a series of Poisson equations, as detailed in the Method section.

\subsection*{Numerical accuracy of the mixed pseudopotential, all-electron approach}

In order to validate our approach for the calculation of the SH parameters using a mixed AE-PP approach, we first consider the Fermi contact term of the $\bm{A}$-tensor for a NV center in a $3\times 3\times 3$ diamond supercell containing 215 atoms, computed using the $\Gamma$-point for Brillouin zone (BZ) sampling. Starting from a case where only the nitrogen atom is treated with an AE description, we gradually increase the number of atoms treated using AE calculations by considering 8 cases (or levels), with 1, 4, 7, 13, 16, 19, 22, and 25 atoms near the defect treated at the AE level. As shown in figure S1 in the Supplementary Information (SI), when 16 neighbor atoms are treated at the AE level, which includes all the C atoms with dangling bonds around the N atom, the value of the Fermi contact is converged (see Table S1 of the SI). In fact, even considering only the nitrogen atom and the three dangling bond carbon atoms at the AE level yields a value of the Fermi contact of -2.125 MHz, which is in very close agreement with the value of -2.096 MHz obtained by a full AE calculation. Our results suggest that by only treating a few atoms at the AE level, the mixed AE-PP calculation, henceforth denoted as FE-mixed, is adequate and accurate to obtain the Fermi contact term. Next, we increased the system size to a $4\times4\times4$ supercell containing 511 atoms, and found that a mixed calculation with the same number of atoms treated at the AE level as in the 215 atom cell, accurately reproduces the Fermi contact term. This indicates that the number of atoms requiring an AE description in a FE-mixed calculation is independent of the system size. We also found that the spin-dipolar term of the $\bm{A}$-tensor (cf. Figure S1 of the SI) is much less sensitive to the number of atoms treated with an AE approach than the Fermi contact.

As previously noted, the evaluation of the $\bm{D}$-tensor is computationally more demanding than that of the $\bm{A}$-tensor, and a complete AE description becomes intractable even for a system with a few hundred atoms. Thus, to validate our mixed approach, we consider NV center in a $2\times 2\times 2$ diamond supercell containing 63 atoms. We performed the mixed calculation with an AE description of the four atoms including the nitrogen atom and the three carbon atoms with dangling bonds. We obtained an excellent agreement for the values of the $\bm{D}$-tensor obtained using FE-mixed and FE-AE calculations. Due to the $C_{3v}$ symmetry of the system, the eigenvalues of the $\bm{D}$-tensor $D_{i}$ ($i\in[1,3]$) follow the relation $D_{1}=D_{2}=-\frac{1}{2}D_{3}$. We report $\frac{3}{2}|D_{3}|$ throughout this manuscript. We obtained a value of 2928.31 MHz in a mixed calculation, to be compared to 2939.47 MHz obtained from a full AE description. Thus, we conclude that the SH parameters obtained for the NV center in diamond by using a FE-mixed approach, where only the nitrogen atom and the three carbon atoms with dangling bonds are described at the AE level, are as accurate as those obtained with a calculation where all atoms are treated with an AE approach. %\bigskip 

We consider next the VV in 4H-SiC (referred to as VV-SiC) \cite{Awshalom-NatureMat2015,Seo-Nature2016}. As in the case of NV-diamond, the VV-SiC has 3 carbon atoms adjacent to a silicon vacancy that contribute to the spin density of the system. Further, VV-SiC also has a silicon atom with 3 dangling bonds,  adjacent to the carbon vacancy that give rise to mid-gap states. Based on our results for NV-diamond,  we only treat these 6 atoms with an AE description. For validation purposes, we first considered a $4\times 4\times 1$ supercell of VV-SiC containing 126 atoms, and we carried out two separate calculations, one using a complete AE description (FE-AE) and the other one using AE descriptions only for the atoms with dangling bonds. We found good agreement for the $\bm{A}$-tensors computed with the two methods (cf. Fig. ~\ref{fig_vv}). Similar to the case of NV-diamond, we conclude that also for VV-SiC, treating a small number of atoms at the AE level (the three carbon and three silicon atoms with dangling bonds) suffices to obtain accurate results. In the case of VV-SiC, a full AE calculation of the $\bm{D}$-tensor is not feasible with reasonable computational resources, and we limited our study to mixed calculations only.   

\begin{figure}[!ht]
  \centering
  \includegraphics[width=0.5\textwidth,trim={0.2cm 0cm 0cm 0},clip]{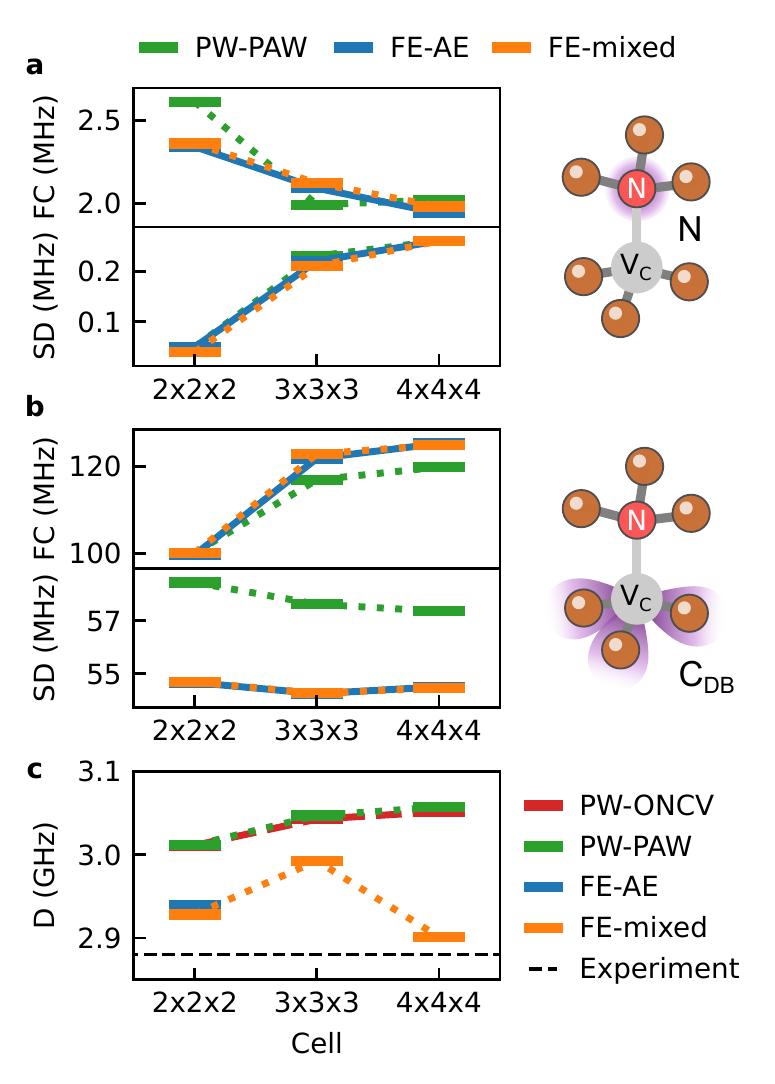}
  \caption{Spin Hamiltonian parameters in NV-diamond. (a) The values of Fermi contact and spin dipolar terms for the nitrogen atom, (b) The values of Fermi contact and spin dipolar terms for the dangling bond carbon atoms. For both types of atoms, the spin dipolar terms are reported in terms of the largest eigenvalue of the $3\times3$ tensor. (c) the zero-field splitting, $\bm{D}$-tensor, where the quantity reported is $\frac{3}{2}|D_{3}|$ with $D_{3}$ being the eigenvalue with largest absolute magnitude. Calculations are performed with finite element (FE) and plane-wave (PW) basis. In the case of FE calculations, we used the proposed mixed all-electron and pseudopotential scheme (FE-mixed), and for select cases we also performed pure all-electron (FE-AE) calculations. All calculations used $\Gamma$-point sampling of the Brillouin zone. Results for higher Brillouin zone sampling are provided in the SI (cf. Table S2).}
  \label{fig_nv}
\end{figure}

\subsection*{Large scale calculations}

We now turn to present results for $\bm{A}$ and $\bm{D}$ obtained with large supercells, using the FE-mixed method. The importance of large supercells to obtain accurate results of spin defects has been emphasized in several recent papers \cite{Davidsson2018,Whiteley2018}. For the NV in diamond, we consider cubic cells ranging from $2\times2\times2$ to $4\times4\times4$ in size, containing 63 to 511 atoms. In the case of VV-SiC, we consider hexagonal cells ranging from $4\times4\times1$ to $8\times8\times2$ in size, containing 126 atoms to 1,022 atoms.

\begin{figure}[!ht]
  \centering
  \includegraphics[width=0.5\textwidth,trim={0.2cm 0cm 0cm 0},clip]{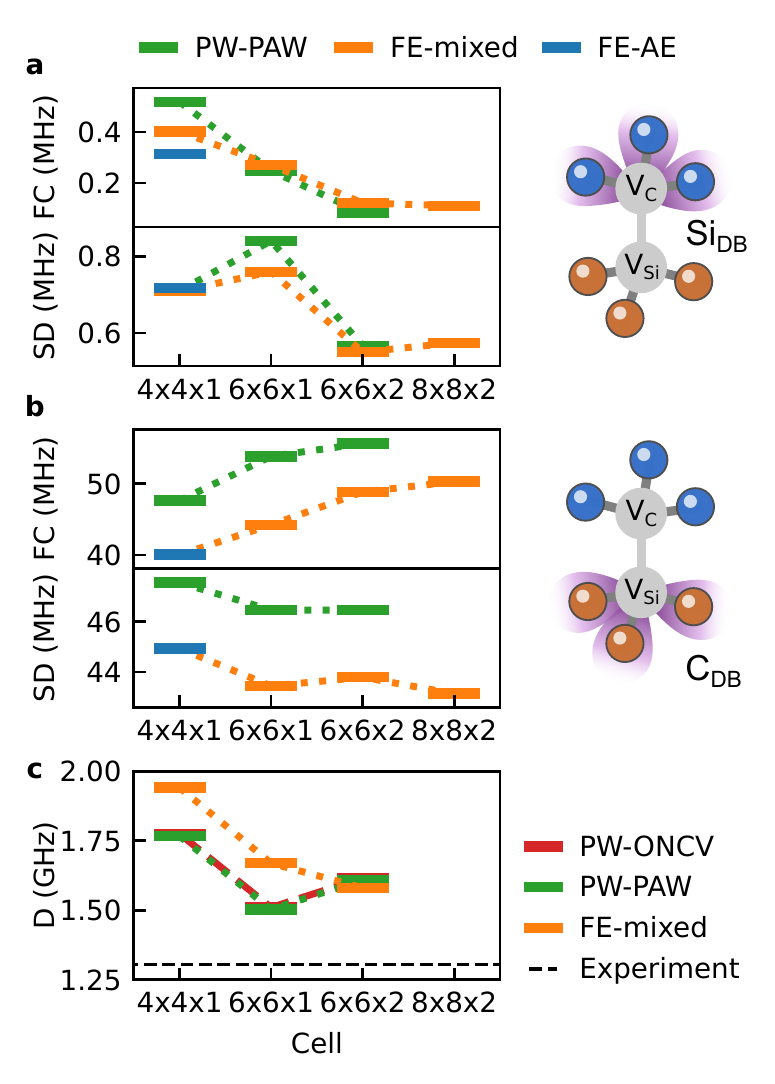}
  \caption{Spin Hamiltonian parameters in VV-SiC. (a) The values of Fermi contact and spin dipolar terms for the dangling bond silicon atoms, (b) The values of Fermi contact and spin dipolar terms for the dangling bond carbon atoms. For both types of atoms, the spin dipolar terms are reported in terms of the largest eigenvalue of the $3\times3$ tensor. (c) the zero-field splitting, $\bm{D}$-tensor, where the quantity reported is $\frac{3}{2}|D_{3}|$ with $D_{3}$ being the eigenvalue with largest absolute magnitude. Calculations are performed with finite element (FE) and plane-wave (PW) basis. PW calculations are carried out using ONCV \cite{Schlipf2015} and PAW  pseudopotentials. In the case of FE calculations, we used the mixed all-electron and pseudopotential scheme (FE-mixed), and for select cases we also performed pure all-electron (FE-AE) calculations. All calculations used $\Gamma$-point sampling of the Brillouin zone. Results for different Brillouin zone sampling are provided in the SI (cf. Table S3). }
  \label{fig_vv}
\end{figure}

The $\bm{A}$-tensor and $\bm{D}$-tensor for the various cell-sizes are shown in Fig.~\ref{fig_nv} and Fig.~\ref{fig_vv}, for NV-diamond and VV-SiC, respectively. For comparison, we also report the same parameters obtained using PW-PP DFT calculations. We found notable differences between the results obtained with the PW-PP method and with the FE-mixed approach, which does yield an overall AE accuracy, as shown above. In the case of VV-SiC (cf. Fig.~\ref{fig_vv}), the errors associated to PW-PP values are as large as $\sim 20\%$ for the value of the Fermi contact term of the $\bm{A}$-tensor. In the case of the $\bm{D}$-tensor, we performed PW calculations using pseudowavefunctions obtained with two PPs: GIPAW \cite{Ceresoli} and ONCV \cite{Schlipf2015}, and found results that agree closely, but differ significantly from those obtained using FE-mixed calculations.     

The ability to compute SH parameters for cells of different sizes provides important insights into the cell-size dependence of the results (see Tables S2 and S3 of the SI). In order to understand the role of defect-defect interactions, we consider the $\bm{A}$-tensor of the dangling bond carbon in NV-diamond computed using $2\times2\times2$ supercells and $3\times3\times3$ k-point sampling with that computed using $3\times3\times3$ supercells and $2\times2\times2$ k-point sampling. These calculations correspond to the same periodic boundary conditions (Born-von Karman boundary condition) for wavefunctions and thus, any change in the $\bm{A}$-tensor is solely due to cell-size effects arising from defect interactions. The Fermi contact value obtained in the two ways changes from 104.74 MHz to 123.36 MHz. Similar cell-size effects arising from the defect-defect interaction are also evident in the case of VV-SiC (cf. Table S3 in the SI). In the case of the $\bm{D}$-tensor, while the cell-size effects are only marginal in  NV-diamond, they are substantial for the VV-SiC (cf. Fig.~\ref{fig_vv}). Our results underscore the importance of carrying out large scale calculations with AE accuracy to obtain accurate SH parameters.

\subsection*{Coherence time in weakly coupled nuclear spin baths: the need for all-electron descriptions}

% We now turn to assess the effect of the accuracy of the computed parameters on the prediction of coherence times.

Having established an efficient approach to compute SH parameters, we now use the spin Hamiltonian to compute dynamical properties of spin defects, in particular coherence times, for which we adopted the cluster correlation expansion (CCE) method \cite{yang2008quantum}. In CCE calculations, the coherence function $L=\frac{\Tr{ \hat \rho(t) \hat S^+}}{\Tr{ \hat \rho(0) \hat S^+}}$ ($\hat \rho$ is the density matrix of the qubit and $\hat S^+$ is the spin raising operator) is approximated as a product of contributions from different nuclear spin clusters. The coherence time is obtained by fitting the coherence function to a compressed exponential function, $L \approx \exp{-(t/T_2)^n}$. The convergence of the results is checked against the size of the spin bath, the number of clusters, and the maximum size of the cluster. Here, the ensemble-averaged coherence function is computed for configurations without nuclear spins with high hyperfine constants ($A_{||} > 1$ MHz). In our CCE calculations, we used hyperfine tensors predicted from DFT calculations for nuclear spins included in the DFT supercell and adopted the point dipole approximation for nuclear spins at distances larger than those included in the supercell.

In general, there are two ways to control nuclear spins in defect systems. The strongly coupled nuclear spins (with hyperfine coupling of order $\sim 1$ MHz) can be directly accessed via radio frequency radiation. Here we define nuclear spins as strongly coupled when their hyperfine parameter is larger than the linewidth of the optically detected magnetic resonance (ODMR) \cite{Christle2015} and separate oscillations in the Ramsey sequence due to the nuclear spin are observed \cite{Bourassa2020}. These nuclear spins can be controlled with short gate times but they are highly susceptible to electron spin induced noise.

The second type of nuclear spins, weakly coupled to the electron spin ($A \ll 1$ MHz) are controlled by dynamical decoupling schemes \cite{Taminiau2014}. Applying refocusing pulses to the central spin may be used to not only increase the coherence time of the defect but also to isolate and control weakly coupled nuclear spins. These spins provide significantly longer coherence times than strongly coupled spins, and the number of weakly coupled nuclear spins is not limited by the short distance to the central spin which is required for strong coupling.

\begin{figure}[!ht]
  \centering
  \includegraphics[width=1\textwidth]{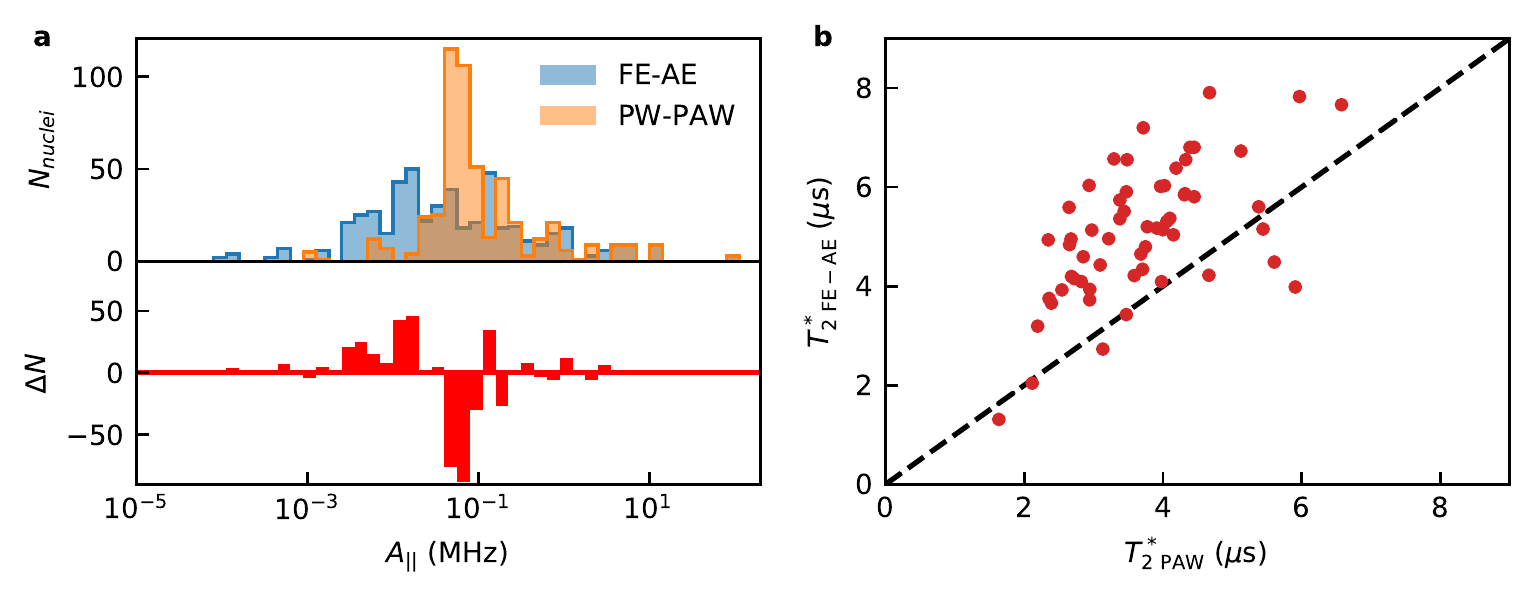}
  \caption{ (a) Histogram of $A_{||}$ (see text) for carbon atoms in a $4\times4\times4$ NV-diamond supercell calculated using PAW and FE-AE methods. (b) Comparison of coherence time $T_2^*$ induced by weakly coupled nuclear spin baths, predicted with hyperfine values from PAW and FE-AE calculations.}
  \label{fig_cce}
\end{figure}

The strength of the nuclear-spin induced dephasing mechanism in spin defects is directly related to the $\bm{A}$-tensor \cite{merkulov2002electron}, which therefore requires accurate calculations. We estimate the sensitivity of the inhomogeneous dephasing time $T_2^*$ on the values of the $\bm{A}$-tensor by carrying out cluster correlation expansion (CCE) calculations, using two sets of hyperfine coupling computed for $4\times4\times4$ supercell of the NV center in diamond: one set for which the $\bm{A}$-tensor is calculated using the PAW reconstructed spin densities, and a second one based on the $\bm{A}$-tensor calculated using FE-AE calculations. Fig. \ref{fig_cce}a shows a histogram of the calculated $A_{||}$ ($A_zz$ in the defect reference frame) obtained using the two methods. We found that for large hyperfine coupling values, the relative difference is rather minor compared to the one for the smaller coupling terms. In order to see the impact of these differences on the dephasing of the electron spin, first we compute the ensemble-averaged dephasing time, $T_2^*$, by considering the decay of the coherence function averaged over a set of nuclear spin configurations. The difference in the ensemble averages dephasing time was found not to be significant (1.35 (FE-AE) $\mu$s vs 1.37 (PAW) $\mu$s). We note that the predicted value is close to the generally accepted value of nuclear-spin-limited $T_2^*$ in diamond of  $\sim$ 1 $\mu$s \cite{barry2020sensitivity}. 

Next, we focus on the single-defect dephasing time in the weakly coupled nuclear spin bath. We select nuclear configurations whose Fourier transform of the free induction decay (FID) contains only one peak. This procedure ensures that the coherence function of the defects in the chosen subset of nuclear configurations does not contain any oscillations due to the nuclear spins with high hyperfine coupling. Hence the procedure guarantees that the nuclear baths contain only weakly coupled nuclear spins and we can thus use an exponential decay fit to obtain the value of $T_2^*$. Figure \ref{fig_cce}b shows the $T_2^*$ for the chosen subset of nuclear configurations. In these systems, the difference between PAW and FE-AE results is significant as the dephasing time differs by a factor of 2 for certain nuclear configurations. We note that in general for weakly coupled spins, the hyperfine coupling computed with PAW are higher (as can be seen in Fig. \ref{fig_cce}a) compared to their FE-AE counterpart, leading to significant differences in the predictions of the dephasing time.

\section*{Discussion} 

In summary, we presented an efficient approach based on real-space density functional theory to compute spin Hamiltonian (SH) parameters. Our approach treats a few selected atoms close to a defect of interest with all-electron (AE) accuracy and uses the pseudopotential (PP) approximation for the remaining atoms; the method uses a finite-element basis set and is systematically convergent with respect to the basis set size. We presented results demonstrating that the accuracy obtained in the computation of the SH parameters using the mixed AE-PP approach is commensurate with that of full AE calculations. The mixed approach in conjunction with the spatial adaptivity of the finite-element basis enabled the computation of hyperfine and zero-field splitting tensors for large systems, containing $\sim$1,000 atoms; these are unprecedented system sizes for the computation of SH parameters with AE accuracy. Our results revealed significant cell-size effects in the calculations of both the hyperfine and the zero-field splitting interaction, indicating that finite size scaling is important in determining accurate values for these tensors. 

We showed that the {\it relative} difference between AE and PP predictions of hyperfine tensors for strongly coupled nuclear spins (those for which $A \ge 1$ MHz) is small.  However, {\it absolute} differences in hyperfine tensors predicted with PW-PP and AE methods for weakly coupled spins ($A \ll 1$ MHz), even when similar in magnitude to those found for strongly coupled spins, may dramatically impact the prediction of $T_2^*$, with differences up to a factor of 2 for certain nuclear spin configurations. We note that, in addition to coherence time calculations, accurate predictions of zero-field splitting and hyperfine tensors for strongly-coupled nuclear spins are important to identify the atomistic structure of novel spin-defects \cite{Ivady2018}; furthermore, accurate predictions of hyperfine tensors for weakly coupled nuclear spins are key for the spatial mapping of experimental multinuclear registers \cite{abobeih2019atomic} and the prediction of plausible memory units in spin centers \cite{Bourassa2020}.

The method introduced here represents a substantial progress towards accurate computations of SH parameters of spin defects in large scale condensed and molecular systems, paving the way to high-throughput screening of novel spin-defects for quantum information technologies.

\section*{Method}

PW based DFT calculations are performed with the \texttt{Quantum ESPRESSO} code \cite{QE} with a kinetic energy cutoff of 75 Ry. GIPAW pseudopotentials developed by Ceresoli~\cite{Ceresoli} were used for the calculation of the $\bm{A}$- and $\bm{D}$-tensor. Moreover, the $\bm{D}$-tensor calculations were also carried out using ONCV pseudopotentials \cite{Schlipf2015}. After solving the Kohn-Sham equations, the $\bm{A}$-tensor is evaluated by PAW reconstruction using the \texttt{GIPAW} code \cite{Bahramy2007}, while the $\bm{D}$-tensor calculations are performed with the \texttt{PyZFS} code~\cite{Ma2020} using normalized pseudo-wavefunctions~\cite{Ivady2014, Falk2014, Seo2017, Whiteley2018}. In $\bm{A}$-tensor calculations, we considered nuclear isotopes $^{13}\text{C}$, $^{14}\text{N}$ and $^{29}\text{Si}$ for C, N and Si, respectively.

FE based DFT calculations are carried out using the \texttt{DFT-FE} code \cite{DFTFE-CPC-2020}. %, which utilizes the FE library \texttt{deal.ii} \cite{dealii}. 
The convergence of the SH parameters with respect to the finite-element mesh discretization parameters is studied on the smallest  supercell for each system considered here. Our convergence study was carried out with respect to the FE polynomial order and the finite-element mesh element size in the vicinity of the AE atoms. Both the polynomial and mesh size requirements are more stringent for the \textbf{A}-tensor calculations compared to the \textbf{D}-tensor ones, as the former depends on the cusps of the spin density at the nucleus. More details related to convergence properties can be found in our previous work~\cite{ghoshPRM2019}.

The computation of the \textbf{A}-tensor uses the self-consistent spin density obtained from the DFT calculation in the following manner. The spin density is evaluated at the FE quadrature points during a self-consistent DFT calculation (as quadrature values are directly used in the ensuing evaluation of integrals), while the Kohn-Sham wavefunctions are computed at the FE nodal points (cf.~\cite{DFTFE-CPC-2020} for details). The Fermi contact term, which requires the value of the spin density at the FE nodes (where the atoms of interest are located), is computed by reevaluating the spin-density at these relevant FE nodes from the Kohn-Sham wavefunctions. The available quadrature point values of the spin density are used for the evaluation of the spin dipolar term, which involves an integral over the spin density (cf. Eq~\eqref{Eq. 3}). However, instead of directly carrying out the integral in Eq.~\ref{Eq. 3}, we evaluate the integral as the Hessian of the potential ($\Phi_s(\bm{r})$) resulting from the spin density, where $\Phi_s(\bm{r})$ is computed as a solution of the partial differential equation (PDE), $\nabla^2 \Phi_s(\bm{r}) = -4\pi n_s(\bm{r})$ with periodic boundary conditions. This reformulation accounts for the contributions of periodic images, which otherwise would have to be explicitly computed in the direct evaluation of the integral in Eq.~\eqref{Eq. 3}. We note that in order to ensure spin neutrality in the solution of the Poisson equation with periodic boundary conditions, a uniform and opposite background with spin density ($-M/\Omega$, where $M$ is the net magnetization of the system, and $\Omega$ is the volume of the supercell) is added to $n_s(\bm{r})$. It is straightforward to show that the contribution of such a uniform background on the Hessian of $\Phi_s(\bm{r})$ is exactly zero.

The calculation of the $\bm{D}$-tensor in real-space is cast into the solution of a series of Poisson equations
\begin{equation} \label{Eq. 4}
    D_{ab} = \frac{1}{2S(2S-1)} \frac{\mu_0}{4\pi} (\gamma_e \hbar)^2 \sum_{i < j}^{occ.} \chi_{ij} ( M^{ij,D}_{ab} -M^{ij,E}_{ab})
\end{equation}
with
\begin{equation} \label{Eq. 5}
    M^{ij,D}_{ab} = \int{\frac{\partial (\phi_{i}(\textbf{r})\phi^{*}_{i}(\textbf{r}))}{\partial r_a}  \Lambda^{jj,D}_{b} (\textbf{r}) d\textbf{r}}\,,\quad \quad
%\end{equation}
%and
%\begin{equation} \label{Eq. 6}
    M^{ij,E}_{ab} = \int{\frac{\partial (\phi_{i}(\textbf{r})\phi^{*}_{j}(\textbf{r}))}{\partial r_a} \Lambda^{ij,E}_{b}(\textbf{r}) d\textbf{r}}\,,
\end{equation}
where $\phi_{i}(\textbf{r})$ is the $i$th single electron wavefunction obtained from DFT, and $\Lambda^{ij}_{b} (\textbf{r})$ is the solution of the Poisson equation $\nabla^2 \Lambda^{ij}_{b} =  -4\pi\frac{\partial ( \phi_{i}^{*}(\textbf{r})\phi_{j}(\textbf{r}))}{\partial r_b}$. The superscript $D$ and $E$ represent the terms corresponding to direct and exchange interactions. $\chi_{ij} = \pm 1$ for $i$th and $j$th wavefunctions having parallel and antiparallel spins, respectively. The most expensive part of the calculation of the $\bm{D}$-tensor  is the solution of the $\frac{N(N+1)}{2}$ Poisson equations, where $N$ is the number of electrons in the system. These Poisson equations are solved using the same FE discretization used to obtain the Kohn-Sham wavefunctions. Hence the reduction in the degrees of freedom achieved via a spatially adaptive FE discretization also aids in improving the computational efficiency of solving the Poisson equations. In addition, Poisson equations corresponding to different pairs of wavefunctions may be solved in parallel since they are independent from each other. 

The PDE involved in the $\bm{D}$-tensor calculation takes the form  $\nabla^2 \Lambda^{ij}_{b} =  -4\pi\frac{\partial ( \phi_{i}^{*}(\textbf{r})\phi_{j}(\textbf{r}))}{\partial r_b}$. Here, the right hand side of the PDE is computed by first evaluating $\phi_{i}^{*}(\textbf{r})\phi_{j}(\textbf{r})$ on the FE nodes and then interpolating the gradient to the quadrature points that are subsequently used in the ensuing integrals of the weak formulation of the finite-element method. The PDEs are solved using the framework for the Poisson problem in the self-consistent cycle of DFT. The potential fields ($\Lambda^{ij,D}_{b}(\textbf{r})$ and $\Lambda^{ij,E}_{b}(\textbf{r})$) obtained from the PDE are defined on the FE nodes and are interpolated to quadrature points before carrying out the integrations in Eq. \ref{Eq. 5}. The interpolation errors systematically decrease with increasing FE polynomial order and decreasing FE mesh size.

As we showed above, the number of \textit{relevant} atoms that need to be treated at the AE level is much smaller ($< 10$) than the total number of atoms in the system, and does not scale with the system size. We note that for systematically convergent calculations, the FE basis functions in AE calculations are $\sim$10-fold larger than in PP calculations for the elements considered here. For the benchmark calculations presented in this work, mixed AE-PP calculations reduce the number of basis functions by $\sim 80\%$ compared to full AE calculations. This advantage of mixed calculations becomes increasingly important for heavier atoms. Importantly, the DFT-FE code has the ability to treat both AE and pseudopotential calculations in the same framework. In DFT-FE, the solution to the Kohn-Sham equations is obtained via the Chebyshev polynomial filtered subspace iteration (ChFSI) procedure~\cite{Zhou-PRE2006,Motamarri2013}. The various computational steps in the ChFSI procedure scales as $\mathcal{O}(MN)$, $\mathcal{O}(MN^2)$ and $\mathcal{O}(N^3)$, where $M$ denotes the number of basis functions, and $N$ is the number of electrons. The computational complexity of solving a Poisson equation is $\mathcal{O}(Mlog(M))$. As the computation of the $\bm{D}$-tensor requires the solution of $\frac{N(N+1)}{2}$ Poisson equations, the computational complexity of the $\bm{D}$-tensor calculations scales as $\mathcal{O}(Mlog(M)N^2)$. Thus the computational cost of both the DFT calculation and the evaluation of the $\bm{D}$-tensor are substantially reduced in mixed AE-PP calculations, given significantly smaller $M$ and $N$.

\section*{Acknowledgements} K.G. and V.G. are grateful for the support of the Department of Energy, Office of Basic Energy Science, through grant number DE-SC0017380, under the auspices of which the computational framework for finite-element based all-electron DFT calculations was developed.  M. O. and G. G. are grateful for the support from AFOSR FA9550-19-1-0358 under which applications to spin-defects were developed and H. M. and G. G. are grateful for the support from MICCoM, under which code development was supported.  MICCoM is part of the Computational Materials Sciences Program funded by the U.S. Department of Energy, Office of Science, Basic Energy Sciences, Materials Sciences and Engineering Division through Argonne National Laboratory, under contract number DE-AC02-06CH11357. This work used computational resources from the University of Michigan through the Greatlakes computing platform, resources from the Research Computing Center at the University of Chicago through UChicago MRSEC (NSF DMR-1420709), and resources of the National Energy Research Scientific Computing Center (NERSC), a U.S. Department of Energy Office of Science User Facility operated under Contract No. DE-AC02-05CH11231.

\section*{Author contributions}
K. G. developed the mixed all-electron pseudopotential approach and performed finite-element calculations. H. M. performed plane-wave calculations. N. O. performed cluster correlation expansion simulations. V.G. and G.G supervised the project. All authors wrote the manuscript.

\bibliographystyle{naturemag}
\bibliography{references}

\clearpage
\setcounter{figure}{0}    
\begin{center}\large\bfseries
Supplementary Information for \\ Spin-spin interactions in solids from mixed all-electron and pseudopotential calculations\textemdash a path to screening materials for spin qubits\end{center}
\renewcommand{\thefigure}{S\arabic{figure}}
\begin{figure}[!h]
\includegraphics[width=3in,trim={0.2cm 0.2cm 0cm 0},clip]{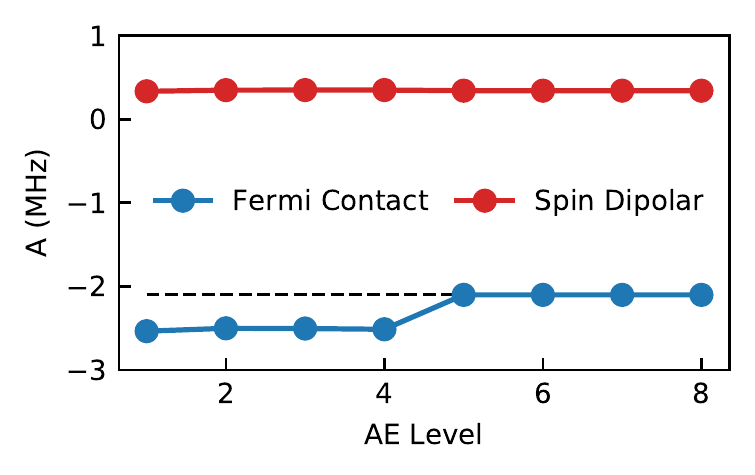}
\caption{ DFT prediction of Fermi contact and spin-dipolar term of the $\bm{A}$-tensor for the nitrogen atom of NV center in $3\times3\times3$ diamond supercell, using a finite-element (FE) basis with a mixed all-electron pseudopotential scheme. FE DFT results converge rapidly with increasing number of all-electron (AE) atoms, denoted by AE level. Eigenvalues with the largest absolute magnitude are shown for spin-dipolar term. %\textcolor{red}{[HM: I suggest we present the AE level shown in Fig. 1 (only N + 3 nearest neighbor C treated with AE) as AE level 2, which demonstrates that using 4 AE atoms yield results almost identical to full AE calculation. The plot shall include AE level 1 (only N), AE level 2 (N + 3C), a few more AE levels with more C, and AE level $\infty$ representing full AE calculation. Then update Table I accordingly.]} 
}
\label{fig3}
\end{figure}

\renewcommand{\thetable}{S\arabic{table}}
\begin{table}[!h]
\caption{Fermi contact and spin dipolar components of the \textbf{A}-tensor for the nitrogen atom of NV center in $3\times3\times3$ diamond supercell, computed using both the mixed calculation as well as pure all-electron calculation (FE-AE). Eigenvalues with the largest absolute magnitude are shown for spin-dipolar term.}
\label{table1}
\begin{tabular}{cccc}
\hline
A (MHz)&        AE Level 4 &   AE Level 5 &               FE-AE \\
\hline
\hline
 Fermi Contact &   -2.513  &    -2.101 & -2.096   \\
%\hline
 Spin Dipolar  &   0.232 &    0.227  &  0.227   \\
\hline
\hline
\end{tabular}
\end{table}

\renewcommand{\thetable}{S\arabic{table}}
\begin{table*}[ht]
\caption{Computed SH parameters of negatively-charged nitrogen-vacancy (NV) center in diamond for various cell-sizes. (a) Fermi-contact and (b) spin-dipolar component of $\bm{A}$-tensor corresponding to the nitrogen atom and dangling bond carbon (DB-C) atom in NV diamond (eigenvalues with the largest absolute magnitude are shown); (c) the zero-field splitting $\bm{D}$-tensor, where the quantity reported is $\frac{3}{2}|D_{3}|$ with $D_{3}$ being the eigenvalue with largest absolute magnitude. Calculations are performed with finite element (FE) and plane-wave (PW) basis. In the case of FE calculations, we used the proposed mixed all-electron and pseudopotential scheme (FE-mixed), and for select cases we also performed pure all-electron (AE) calculations. }\label{tab:table2}
\addtocounter{table}{-1}
\renewcommand{\thetable}{S\arabic{table}.a}
\caption {Fermi Contact (in MHz)} \label{tab:NV FC}
%\begin{subtable}
%\caption{A subtable}
\begin{tabular}{c|c|c|c|c|c|c}
\hline
Atom & System size &   Number of atoms &    BZ sampling  & PW &                   FE-mixed &                             FE-AE \\
\hline
\hline
    &                &      & $\Gamma$          & -2.605 & -2.360 & -2.344 \\
    &$2\times2\times2$ &  63  & $2\times2\times2$ & -1.967 & -1.692 & - \\
    &                &      & $3\times3\times3$ & -1.930 & -1.733 & - \\\cline{2-7}
%\hline
    &                &      & $\Gamma$           & -2.180 & -2.125 & -2.095 \\
N    &$3\times3\times3$ & 215  & $2\times2\times2$  & -1.993 & -1.939 & -      \\\cline{2-7}
%\hline
    &$4\times4\times4$ & 511  & $\Gamma$           & -2.021 & -1.986 &   -1.949    \\
\hline
    &               &      & $\Gamma$          & 100.114 & 100.077 & 99.658 \\
    &$2\times2\times2$ &  63  & $2\times2\times2$ & 103.651 & 105.494 & - \\
    &                &      & $3\times3\times3$ & 103.932 & 104.774 & - \\\cline{2-7}
%\hline
    &                &      & $\Gamma$           & 116.891 & 122.813 & 121.809 \\
DB-C    &$3\times3\times3$ & 215  & $2\times2\times2$  & 117.293 & 123.360 & -      \\\cline{2-7}
%\hline
    &$4\times4\times4$ & 511  & $\Gamma$           & 119.809 & 124.927 &   125.228    \\
\hline
\hline
\end{tabular} 
%\end{subtable}
\addtocounter{table}{-1}
\renewcommand{\thetable}{S\arabic{table}.b}
\caption {Spin Dipolar (in MHz)} \label{tab:NV SD}

\begin{tabular}{c|c|c|c|c|c|c} 
\hline
Atom & System size &   Number of atoms &     BZ sampling & PW &                   FE-mixed &                             FE-AE \\
\hline
\hline
    &                &      &  $\Gamma$          &  -0.052 & -0.049 & -0.054    \\
    & $2\times2\times2$ &  63  &  $2\times2\times2$ &  0.416 & 0.349 & -  \\
    &                &      &  $3\times3\times3$ &  0.346 & 0.353 & -   \\\cline{2-7}
%\hline
    &                &      &  $\Gamma$          &  0.233 & 0.219 & 0.227   \\
  N & $3\times3\times3$ &  215 &  $2\times2\times2$ &  0.273 & 0.255 & -       \\\cline{2-7}
%\hline
    & $4\times4\times4$ &   511 &  $\Gamma$ &  0.270 &      0.262 & 0.261    \\
\hline
    &                &      &  $\Gamma$          &  58.451 & 54.674 & 54.657    \\
    & $2\times2\times2$ &  63  &  $2\times2\times2$ &  57.898 & 54.366 & -  \\
    &                &      &  $3\times3\times3$ &  57.914 & 54.332 & -   \\\cline{2-7}
%\hline
    &                &      &  $\Gamma$          &  57.628 & 54.254 & 54.235   \\
DB-C& $3\times3\times3$ &  215 &  $2\times2\times2$ &  57.394 & 54.054 & -       \\\cline{2-7}
%\hline
    & $4\times4\times4$ &   511 &  $\Gamma$ &  57.360 &      54.450 & 54.496    \\
\hline
\hline
\end{tabular}
\addtocounter{table}{-1}
\renewcommand{\thetable}{S\arabic{table}.c}
\caption {\textbf{D}-tensor (in MHz)} \label{tab:NV Dtensor}
\begin{tabular}{c|c|c|c|c|c}
\hline
System size &   Number of atoms &     PW-GIPAW &       PW-ONCV &            FE-mixed &                             FE-AE \\
\hline
\hline
  $2\times2\times2$ &   63  &   3011.47 &   3011.13 &    2928.31 &        2939.47    \\
%\hline
  $3\times3\times3$ &   215 &   3046.91 &   3042.61 &    2992.58 &          -   \\
%\hline
  $4\times4\times4$ &   511 &   3057.04 &   3051.45 &    2901.28 &          -    \\
\hline
\hline
\end{tabular}
\label{hyperfine_dip_table2}
\end{table*}
 
\begin{table*}[ht]
\caption{Computed SH parameters of divacancy (VV) in SiC for various cell-sizes. (a) Fermi-contact and (b) spin-dipolar component of $\bm{A}$-tensor corresponding to the dangling bond carbon (DB-C) and dangling bond silicon (DB-Si) atom in VV-SiC (eigenvalues with the largest absolute magnitude are shown); (c) the zero-field splitting $\bm{D}$-tensor, where the quantity reported is $\frac{3}{2}|D_{3}|$ with $D_{3}$ being the eigenvalue with largest absolute magnitude. Calculations are performed with finite element (FE) and plane-wave (PW) basis. In the case of FE calculations, we used the proposed mixed all-electron and pseudopotential scheme (FE-mixed), and for select cases we also performed pure all-electron (AE) calculations. 
}\label{tab:table3}
\addtocounter{table}{-1}
\renewcommand{\thetable}{S\arabic{table}.a}
\caption {Fermi Contact (in MHz)} \label{tab:SiC FC}
%\begin{subtable}
%\caption{A subtable}
\begin{tabular}{c|c|c|c|c|c|c}
\hline
Atom & System size &   Number of atoms & BZ Sampling & PW & FE-mixed & FE-AE \\
\hline
\hline
   &                 &       & $\Gamma$          & 47.612 & 40.046 &  40.100 \\
   &$4\times4\times1$ &  126  & $2\times2\times2$ & 50.994 & 43.999 &  - \\
   &                    &       & $3\times3\times2$ &      - & 43.794  &  - \\\cline{2-7}
%\hline
   &                &       & $\Gamma$          & 53.822 & 44.305 &  - \\
  DB-C &$6\times6\times1$ &  286  & $2\times2\times2$ & 54.568 & 45.261    &  - \\\cline{2-7}
%\hline
   &$6\times6\times2$ &  574  & $\Gamma$          & 55.608 & 48.831 &  -    \\\cline{2-7}
%\hline
   &$8\times8\times2$ &  1022  & $\Gamma$          &   -    & 50.272 &  -    \\
 \hline
   &                &       & $\Gamma$          & 0.518 & 0.402 &  0.315 \\
   &$4\times4\times1$ &  126  & $2\times2\times2$ & 0.419 & 0.093 &  - \\
   &                    &       & $3\times3\times2$ &      - & 0.081  &  - \\\cline{2-7}
%\hline
   &                &       & $\Gamma$          & 0.248 & 0.271 &  - \\
 DB-Si  &$6\times6\times1$ &  286  & $2\times2\times2$ & 0.320 & 0.118    &  - \\\cline{2-7}
%\hline
   &$6\times6\times2$ &  574  & $\Gamma$          & 0.082 & 0.122 &  -    \\\cline{2-7}
%\hline
   &$8\times8\times2$ &  1022  & $\Gamma$          &   -    & 0.111 &  -    \\
\hline
\hline
\end{tabular} 
%\end{subtable}
\addtocounter{table}{-1}
\renewcommand{\thetable}{S\arabic{table}.b}
\caption {Spin Dipolar (in MHz)} \label{tab:SiC SD}
\begin{tabular}{c|c|c|c|c|c|c} 
\hline
Atom& System size &   Number of atoms &     BZ Sampling  & PW &                   FE-mixed &                             FE-AE \\
\hline
\hline
  &                 &       & $\Gamma$          & 47.540 & 44.937 & 44.928 \\
  &$4\times4\times1$ &  126  & $2\times2\times2$ & 46.728 & 44.455 &  - \\
  &                  &       & $3\times3\times2$ &   -    & 44.189  &  - \\\cline{2-7}
%\hline
  &                  &       & $\Gamma$          & 46.452 & 43.449 &  - \\
DB-C  &$6\times6\times1$ &   286 & $2\times2\times2$ & 46.373 & 43.506 &  - \\\cline{2-7}
%\hline
  &$6\times6\times2$ &   574 & $\Gamma$          & 46.449 & 43.793  &  -    \\\cline{2-7}
%\hline
  &$8\times8\times2$ &   1022 & $\Gamma$           &  46.345     & 43.144  &  -    \\
\hline
  &                  &       & $\Gamma$          & 0.710 & 0.708 & 0.717 \\
  &$4\times4\times1$ &  126  & $2\times2\times2$ & 0.574 & 0.571 &  - \\
  &                  &       & $3\times3\times2$ &   -    &0.569  &  - \\\cline{2-7}
%\hline
  &                  &       & $\Gamma$          & 0.839 & 0.759 &  - \\
DB-Si  &$6\times6\times1$ &   286 & $2\times2\times2$ & 0.592 & 0.538 &  - \\\cline{2-7}
%\hline
  &$6\times6\times2$ &   574 & $\Gamma$          & 0.567 & 0.550  &  -    \\\cline{2-7}
%\hline
 &$8\times8\times2$ &   1022 & $\Gamma$           &  0.573     & 0.574  &  -    \\
\hline
\end{tabular}
\addtocounter{table}{-1}
\renewcommand{\thetable}{S\arabic{table}.c}
\caption {\textbf{D}-tensor (in MHz)} \label{tab:SiC Dtensor}
\begin{tabular}{c|c|c|c|c}
\hline
System size &   Number of atoms &     PW &       ONCV &            FE-mixed \\
\hline
\hline
  $4\times4\times1$ &   126  &   1766.900 &       1775.320 &        1941.214    \\
%\hline
  $6\times6\times1$ &   286 &    1503.270 &       1510.340 &         1669.392    \\
%\hline
  $6\times6\times2$ &   574 &    1608.980 &       1615.900 &          1580.365     \\
\hline
\hline
\end{tabular}

\label{hyperfine_dip_table3}
\end{table*}

\end{document}